\documentclass[amsmath,amssymb,floatfix,prl]{revtex4}

\usepackage{graphicx}
\begin{document}

\title{Revisiting  a variational study of the quantum $J_{1}-J_{1}^{'}-J_{2}$ spin-1/2 Heisenberg antiferromagnet}
\author{Kevin J.  Urcia Vidarte}
\affiliation{Universidad Nacional Pedro Ruiz Gallo, Escuela de F\'{\i}sica, Av. Juan XXIII 391, Lambayeque, Per\'{u}.}

\author{Luigui A. Chero Nunura}
\affiliation{Universidad Nacional Pedro Ruiz Gallo, Escuela de F\'{\i}sica, Av. Juan XXIII 391, Lambayeque, Per\'{u}.}

\author{Octavio D. Rodriguez Salmon \footnote{Corresponding author.\\ E-mail address: octaviors@gmail.com}}
\affiliation{Universidade Federal do Amazonas, Departamento de F\'{\i}sica, 3000, Japiim, 69077-000, Manaus-AM, Brazil}

\author{Jos\'{e} Ricardo de Sousa}
\affiliation{Departamento de F\'{\i}sica, Universidade Federal do Amazonas, 3000, Japiim,
69077-000, Manaus-AM, Brazil}
\affiliation{National Institute of Science and Technology for Complex Systems, 3000, Japiim,
69077-000, Manaus-AM, Brazil}

\begin{abstract}
The variational study of the  ground state of the spin$-1/2$ anisotropic Heisenberg antiferromagnet has been revisited on a square lattice by improving and correcting past numerical results found in Sol. State. Comm. 165, 33 (2013).   The  Hamiltonian has been implemented on a square lattice with antiferromagnetic interactions between nearest- and next-nearest  neighbors. The nearest-neighbor couplings have different strengths, namely, $J_{1}$ and $J_{1}^{'}$, for the x and y directions, respectively. These couplings compete with the next-nearest ones denoted by $J_{2}$. We obtained a new phase diagram in the $\lambda - \alpha$ plane, where $\lambda= J_{1}^{'}/J_{1}$ and $\alpha = J_{2}/J_{1}$, whose topology is slightly different of that previously found. There is no direct frontier dividing the collinear (CAF) and the antiferromagnetic order (AF), rather,  the quantum paramagnetic phase (QP) separates these two phases for all positive values of $\lambda$ and $\alpha$. The true nature of the frontiers has been obtained by  scanning rigorously the relevant points of  the $\lambda-\alpha$ plane.

\end{abstract}

\maketitle

\section{Introduction}

Frustration is an interesting phenomenom in magnetism where a spin is unable to find an orientation that satisfies all its exchange interactions with its neighboring spins \cite{diep}. The lattice structure or  competing interactons are the causes of this phenomenom. The ground state of the $J_{1}-J_{2}$ Heisenberg model is a good example of this behavior due its competing interactions. Experimental realizations in vanadium phosphates compounds, such as $\bf{VOMoO_{4}}$, $\bf{Li_{2}VOSiO_{4}}$ and $\bf{Li_{2}VOGeO_{4}}$ \cite{melzi, carreta, carreta2, rosner, bombardi}, prove that there exist  prototypes in nature of the two-dimensional frustrated quantum Heisenberg antiferromagnet. For instance, isostructural compounds $\bf{Li_{2}VOSiO_{4}}$ and $\bf{Li_{2}VOGeO_{4}}$, formed by layers of $V^{+4}$ ions of $S=1/2$ on a square latice, show evidences of  a collinear order  for  $\alpha = J_{2}/J_{1}  > 0.5$, where $J_{1}$ and $J_{2}$ are the strength of the nearest- and next-nearest-neighbor couplings, respectively. This is in agreement with theoretical predictions \cite{darradi} that establish two long-range magnetic orders, the antiferromagnetic (AF) and the collinear one (CAF), and an intermediate disordered phase (QP), for $0.4 < \alpha < 0.6$, whose quantum properties are not fully understood yet. An illustration of how we could imagine the spins in the possible ordered phases is shown in Fig.1.
\vskip \baselineskip 
Accordingly,  the manipulation of the value of $\alpha$ enables the exploration of the phases of the ground state, and it can also be possible  experimentaly when applying high pressures causing significant contractions on the $\bf{Li-O}$ bonds of the $\bf{Li_{2}VOSiO_{4}}$ compound \cite{pavarini}. These physical realizations  show the relevancy of the spin$-1/2$ $J_{1}-J_{2}$ Heisenberg model in the square lattice, whose ground state is  a good candidate for the spin liquid state. According to Anderson \cite{anderson},  low spin, low spatial dimension, and high frustration can lead to this phase, and the  $J_{1}-J_{2}$ model meets these requirements. 
\vskip \baselineskip  
On the other hand, it is important to consider the  $J_{1}-J_{1}^{'}-J_{2}$ spin-1/2 Heisenberg antiferromagnet, which   is a generalization of the spin$-1/2$ $J_{1}-J_{2}$ Heisenberg model on the square lattice. The model was firstly studied by  Nersesyan and Tsvelik \cite{nersesyan}, then other researchers have treated it \cite{starykh,sindzingre,igarashi,bishop,jiang}. By definition,  not all the nearest-neighbor interactions are equivalent in the $J_{1}-J_{1}^{'}-J_{2}$  model, this is why the horizontal and  vertical  couplings are denoted by $J_{1}$ and $J_{1}^{'}$, respectively.  As in the original model,  the frustration parameter is given by $\alpha = J_{2}/J_{1}$, but the anisotropy between the vertical and the horizonal nearest-neighbor interactions induces   the introduction of the parameter $\lambda = J_{1}^{'}/J_{1}$, where $\lambda \neq 1$. The introduction of this anisotropy parameter is not only for theoretical interest, there exist compounds whose couplings show  in fact that $J_{1}^{'}$ and $J_{1}$ can have different values. For instance, in $\bf{SrZnVO(PO_{4})_{2}}$ was found that $J_{1}^{'}/J_{1} = 0.7$ and $J_{2}^{'}/J_{2} = 0.4$ \cite{tsirlin}. To illustrate it,  we depicted in Fig.2 the magnetic interactions in the $\bf{[VOPO_{4}]}$ layers showing that even  the NNN couplings may have different values according to the spatial direction, so it introduces two kind of next-nearest neighbor interactions, namely, $J_{2}^{'}$ and $J_{2}$. Nevertheless,  in this work we consider equal next-nearest neighbors, so  $J_{2}^{'}=J_{2}$.
\vskip \baselineskip
 The aim of this paper is to show the correct numerical results of the varional study of the ground state of the  $J_{1}-J_{1}^{'}-J_{2}$ model studied by  Mabelini et al. \cite{orlando}, whose work did not obtain the precise topology of the phase diagram in that approach. We have also calculated numerically the energy surface  in  the $\lambda - \alpha$ plane that ensures us which phase minimizes the energy of the Hamiltonian for a given region in the $\lambda - \alpha$ plane. The rest of this work is organized as follows: In Section 2, the Hamiltonian is presented and treated by a variational method in the mean-field approximation. In Section 3, the main results are shown and discussed. Finally, the conclusions are given in Section 4. 

\section{The variational study of the model}

The Hamiltonian describing the  $J_{1}-J'_{1}-J_{2}$ model in the square lattice is given by:
\begin{equation}
\hat H = J_{1}\sum_{(ij)x}\vec{\sigma_{i}}\cdot \vec\sigma_{i+\vec\delta x}+J'_{1}\sum_{(ij)y}\vec{\sigma_{i}}\cdot \vec\sigma_{i+\vec\delta y} + J_{2}\sum_{<ij>}\vec{\sigma_{i}}\cdot \vec{\sigma_{j}},
\label{hamj1j2}
\end{equation}
where the first and the second  sum correspond to nearest-neighbor interactions along the x and the y axis, respectively, and third sum is for the next-nearest-neighbor interactions. 
\vskip \baselineskip

In Fig. 2 we show how the $J_{1}-J'_{1}-J_{2}$ model is implemented in the square lattice. The idea was originally developed by Oliveira \cite{oliveira}, which consists of considering a trial ground-state wave vector proposed as the product of the plaquettes $\lbrace |\phi_{0l}\rangle\rbrace$, given by :
\begin{eqnarray}
        \nonumber |\Psi_{0}\rangle&=& \prod_{l=1}^{N/4} \; |\phi_{0l}\rangle\\
         &=& |\phi_{01}\rangle\otimes|\phi_{02}\rangle\otimes \cdots \otimes|\phi_{0N/4}\rangle ,
        \end{eqnarray}
where the wave vector of the plaquette $l$ is the linear combination of the  states of the spins 1, 2, 3 and 4 corresponding to the plaquette $A$ used as a reference (see Fig.2). So,  $|\phi_{0l}\rangle$ is the superposition of the independent states $\lbrace|n\rangle\rbrace$ with total spin projection $\sigma^{z}=0$, where the coefficients  $\lbrace a_{n} \rbrace$ are the variational parameters restricted to the normalization condition $\displaystyle \sum_{n=1}^{6}a_{n}^{2}=1$. In mathematical terms, each plaquette state is expressed as

\begin{equation}
|\phi_{0l}\rangle=\sum_{n=1}^{6}a_{n}|n\rangle
\end{equation}
where $|1\rangle=|+-+-\rangle$, $|2\rangle=|-+-+\rangle$, $|3\rangle=|++--\rangle$, $|4\rangle=|-++-\rangle$, $|5\rangle=|--++\rangle$, $|6\rangle=|+--+\rangle$. However, it is algebraically convenient to use the variables  $x$, $y$, $z$, $u$, $v$ y $w$, such that $x=(a_{1}+a_{2})/\sqrt{2}$, $y=(a_{3}+a_{5})/\sqrt{2}$, $z=(a_{4}+a_{6})/\sqrt{2}$, $u=(a_{1}-a_{2})/\sqrt{2}$, $v=(a_{3}-a_{5})/\sqrt{2}$, $w=(a_{4}-a_{6})/\sqrt{2}$. In this way, the normalization condition for the coefficients $\lbrace a_{n} \rbrace$ is now written as $x^{2}+y^{2}+z^{2}+u^{2}+v^{2}+w^{2}=1$. \\\\
 The  magnetization of each site  of a plaquette is the mean of the spin operator for that site with respect to the vector state of that plaquette. For instance, the magnetization of the spin 1 of the plaquette A is given by 

\begin{align}
\langle\vec{\sigma}_{1}\rangle&=\langle\phi|\vec{\sigma}_{1}|\phi\rangle=\sum_{n=1}^{6}a_{n}^{2}\langle n|\vec{\sigma}_{1}|n\rangle\\
&=\langle\phi|\sigma_{1}^{z}|\phi\rangle \hat{z} = (a_{1}^{2}-a_{2}^{2}+a_{3}^{2}-a_{4}^{2}-a_{5}^{2}+a_{6}^{2})\hat{z}
\end{align}

Thus, the magnetization $m_{1}$ of the site 1 in the plaquette A is
\begin{equation}
m_{1}=a_{1}^{2}-a_{2}^{2}+a_{3}^{2}-a_{4}^{2}-a_{5}^{2}+a_{6}^{2}.
\end{equation}

In the same way we can determine the other magnetizations, so, the four ones can be written in terms of the variables  $x$, $y$, $z$, $u$, $v$ y $w$, as

\begin{equation}
\begin{split}
m_{1}&=2(xu+yv-zw)\\
m_{2}&=2(-xu+yv+zw)\\
m_{3}&=2(xu-yv+zw)\\
m_{4}&=2(-xu-yv-zw)
\end{split}
\label{mag}
\end{equation}
\vskip \baselineskip
Let $E_{0}$ be the energy per spin of the ground state in $J_{1}$ units, so it is calculated by computing the mean value of the Hamiltonian operator in the trial wave function. Accordingly, $E_{0}=\langle\Psi_{0}|H|\Psi_{0}\rangle/4NJ_{1}$. We can split the calculation of $E_{0}$ into  two parts as  $E_{0} = \frac{1}{4} (\epsilon_{A}+\epsilon_{A}^{int})$, where

\begin{equation}
\begin{split}
	\epsilon_{A}=  \langle\vec{\sigma}_{1}\cdot\vec{\sigma}_{2}\rangle_{A}+\langle\vec{\sigma}_{3}\cdot\vec{\sigma}_{4}\rangle_{A}+\lambda[\langle\vec{\sigma}_{2}\cdot\vec{\sigma}_{3}\rangle_{A}+\langle\vec{\sigma}_{1}\cdot\vec{\sigma}_{4}\rangle_{A}]+\quad\quad\quad\quad\quad\\
	\quad\quad\quad\alpha[\langle\vec{\sigma}_{1}\cdot\vec{\sigma}_{3}\rangle_{A}+\langle\vec{\sigma}_{2}\cdot\vec{\sigma}_{4}\rangle_{A}]
\end{split}
\end{equation}
y
\begin{equation}
\begin{split}
\epsilon_{A}^{int}&=  \langle\vec{\sigma}_{2}\rangle_{A}\cdot\langle\vec{\sigma}_{8}\rangle_{D}+\langle\vec{\sigma}_{3}\rangle_{A}\cdot\langle\vec{\sigma}_{9}\rangle_{C}+\lambda[\langle\vec{\sigma}_{1}\rangle_{A}\cdot\langle\vec{\sigma}_{5}\rangle_{B}+\langle\vec{\sigma}_{2}\rangle_{A}\cdot\langle\vec{\sigma}_{6}\rangle_{B}]+\\
                  &\quad\alpha[\langle\vec{\sigma}_{1}\rangle_{A}\cdot\langle\vec{\sigma}_{6}\rangle_{B}+\langle\vec{\sigma}_{2}\rangle_{A}\cdot\langle\vec{\sigma}_{5}\rangle_{B}+\langle\vec{\sigma}_{2}\rangle_{A}\cdot\langle\vec{\sigma}_{7}\rangle_{D}+\langle\vec{\sigma}_{6}\rangle_{B}\cdot\langle\vec{\sigma}_{8}\rangle_{D}+\\
                  &\quad\langle\vec{\sigma}_{3}\rangle_{A}\cdot\langle\vec{\sigma}_{8}\rangle_{D}+\langle\vec{\sigma}_{2}\rangle_{A}\cdot\langle\vec{\sigma}_{9}\rangle_{D}]
\end{split}
\end{equation}

In order to perform the  calculation of the brackets we use the properties of the Pauli operators, such that $\sigma^{x}|\pm\rangle=|\mp\rangle$, $\sigma^{y}|\pm\rangle=\pm i|\mp\rangle$, $\sigma^{z}|\pm\rangle=\pm|\pm\rangle$, and the fact that $|\phi_{0A}\rangle=a_{1}|1\rangle+a_{2}|2\rangle+a_{3}|3\rangle+a_{4}|4\rangle+a_{5}|5\rangle+a_{6}|6\rangle$, so 
\begin{multline*}
\epsilon_{A}=4(a_{1}+a_{2})(a_{4}+a_{6})+2(-a_{1}^{2}-a_{2}^{2}+a_{3}^{2}-a_{4}^{2}+a_{5}^{2}-a_{6}^{2})+\\
\lambda[4(a_{1}+a_{2})(a_{3}+a_{5})+2(-a_{1}^{2}-a_{2}^{2}-a_{3}^{2}+a_{4}^{2}-a_{5}^{2}+a_{6}^{2})]+\\
\alpha[4(a_{3}+a_{5})(a_{4}+a_{6})+2(a_{1}^{2}+a_{2}^{2}-a_{3}^{2}-a_{4}^{2}-a_{5}^{2}-a_{6}^{2})].
\end{multline*}

Now, the above expression of the energy $\epsilon_{A}$ can be written through the variables $x$, $y$, $z$, $u$, $v$ and $w$, as follows:

\begin{equation}
\begin{split}
\epsilon_{A}=-2(\lambda+1)(x^{2}+u^{2})+2(1-\lambda)[(y^{2}+v^{2})-(z^{2}+w^{2})]+\quad\quad\quad\quad\quad\quad\\
             \quad\quad\quad8x(\lambda y+z)+\alpha[2-4(y-z)^{2}-4v^{2}-4w^{2}]
\end{split}
\end{equation}

By considering the frontier conditions  $\vec{\sigma}_{5} = \vec{\sigma}_{4}$, $\vec{\sigma}_{6} = \vec{\sigma}_{3}$, $\vec{\sigma}_{8} = \vec{\sigma}_{1}$, $\vec{\sigma}_{9} = \vec{\sigma}_{4}$, $\vec{\sigma}_{7} = \vec{\sigma}_{4}$, and using the magnetization of the sites of the pĺaquette A given in Eq.(\ref{mag}), we can express $\epsilon_{A}^{int}$ as 
\begin{equation}
\epsilon_{A}^{int}=-8(\lambda+1)x^{2}u^{2}+8(1-\lambda)(y^{2}v^{2}-z^{2}w^{2})+24\alpha(x^{2}u^{2}-y^{2}v^{2}-z^{2}w^{2})
\end{equation}

Therefore, the energy per spin $E_{0}$ is finally written as 
\begin{equation}
\begin{split}
E_{0}=-\dfrac{(\lambda+1)}{2}(x^{2}+u^{2})-2(\lambda+1)x^{2}u^{2}+\dfrac{(1-\lambda)}{2}[(y^{2}+v^{2})-(z^{2}+w^{2})]+2x(\lambda y+z)+\quad\quad\quad\quad\quad\quad\quad\quad\quad\quad\quad\quad\quad\quad\quad\quad\quad\quad\quad\quad\quad\\
       2(1-\lambda)(y^{2}v^{2}-z^{2}w^{2})+\alpha[\dfrac{1}{2}-(y-z)^{2}-v^{2}-w^{2}+6x^{2}u^{2}-6y^{2}v^{2}-6z^{2}w^{2}]\quad\quad\quad\quad\quad\quad\quad\quad\quad\quad\quad\quad\quad\quad\quad\quad\quad\quad\quad\quad\quad\quad\quad
\end{split}
\end{equation}

On the other hand, the functional that minimizes $E_{0}$ subjected to the normalization condition is given by 
\begin{equation}
\mathcal{F}(x,y,z,u,v,w,\eta)=\epsilon_{0}-\eta(x^{2}+y^{2}+z^{2}+u^{2}+v^{2}+w^{2}-1),
\end{equation}

where $\eta$ is a lagrange multiplier. Thus, the extremization $\delta\mathcal{F}=0$ leads to the following set of nonlinear equations:

\begin{equation}
\left\lbrace \begin{array}{lcc}  
             \begin{split}          
            -(\lambda+1)x-4(\lambda+1)xu^{2}+2(\lambda y+z)+12\alpha xu^{2} &=2\eta x\\
            (1-\lambda)y+4(1-\lambda)yv^{2}+2\lambda x-2\alpha(y-z)-12\alpha yv^{2}&=2\eta y\\
            -(1-\lambda)z-4(1-\lambda)zw^{2}+2x+2\alpha(y-z)-12\alpha zw^{2}&=2\eta z\\
            -(\lambda+1)u-4(\lambda+1)x^{2}u+12\alpha x^{2}u&=2\eta u\\
            (1-\lambda)v+4(1-\lambda)y^{2}v-2\alpha v(1+6y^{2})&=2\eta v\\
            -(1-\lambda)w-4(1-\lambda)z^{2}w-2\alpha w(1+6z^{2})&=2\eta w
            \end{split}             
            \end{array}
\right.           
\end{equation}

These equations were not rightly written in reference \cite{orlando}. We may note that in the isotropic case  $\lambda=1$, we recover the same equations  obtained by Oliveira for the $J_{1}-J_{2}$ model \cite{oliveira}. In order to find the variational parameters for each phase of the ground state, we have to impose the corresponding configurations for the magnetizations of each spin, so we will analyze the three possible phases: 
\begin{enumerate}
\item Quantum Paramagnetic Phase (PQ): $m_{1}=m_{2}=m_{3}=m_{4}=0$, which implies that $xu=yv=zw=0$. Thus, it generates the following set of equations: 
\begin{align*}
-(\lambda+1)x+2(\lambda y+z)&=2\eta x\quad\quad\quad\\
(1-\lambda)y+2\lambda x-2\alpha(y-z)&=2\eta y\\
-(1-\lambda)z+2x+2\alpha(y-z)&=2\eta z\\
-(\lambda+1)u&=2\eta u\\
(1-\lambda)v-2\alpha v&=2\eta v\\
-(1-\lambda)w-2\alpha w&=2\eta w
\end{align*}

\item Antiferromagnetic phase (AF): $m_{1}=-m_{2}=m_{3}=-m_{4}$, which leads to $v=w=0$ y $u\neq0$, obtaining:

\begin{align*}
-(\lambda+1)x-4(\lambda+1)xu^{2}+2(\lambda y+z)+12\alpha xu^{2}&=2\eta x\quad\quad\quad\quad\\
(1-\lambda)y+2\lambda x-2\alpha(y-z)&=2\eta y\\
-(1-\lambda)z+2x+2\alpha(y-z)&=2\eta z\\
-(\lambda+1)u-4(\alpha+1)x^{2}u+12\alpha x^{2}u&=2\eta u
\end{align*}

\item Collinear phase (CAF):$m_{1}=m_{2}=−m_{3}=−m_{4}$, then we have two cases $u=w=0$ and $v\neq0$, or $u=v=0$ and $w\neq0$. The latter case produces the following set of equations:
\begin{align*}
-(\lambda+1)x+2(\lambda y+z)&=2\eta x\quad\quad\quad\quad\quad\quad\\
(1-\lambda)y+2\lambda x-2\alpha(y-z)&=2\eta y\\
-(1-\lambda)z-4(1-\lambda)zw^{2}+2x+2\alpha(y-z)-12\alpha zw^{2}&=2\eta z\\
-(1-\lambda)w-4(1-\lambda)z^{2}w-2\alpha w(1+6z^{2})&=2\eta w
\end{align*}
\end{enumerate}

These equations will help us to find the zone in the $\lambda-\alpha$ plane where $E_{0}$ is minimized, and this determines the corresponding stable phase. On the other hand, the order parameters $m_{\textup{AF}} = (m_{1}-m_{2}+m_{3}-m_{4})/4$ and $m_{\textup{CAF}}=(m_{1}+m_{2}-m_{3}-m_{4})/4$ can be  calculated as functions of the parameter $\alpha$ for a given value of the parameter $\lambda$. In the next section we analyze the results based on these formulations. 

\section{Results}

In Fig.4 we show the order parameters $m_{\textup{AF}}$  and $m_{\textup{CAF}}$ as functions of the frustation parameter $\alpha$, for different values of  $\lambda$. We can observe that the curves $m_{\textup{AF}}$ (which are on the left hand) fall to zero for certain values of $\alpha$ denoted by $\alpha_{1c}$. The curves on the right hand of this figure correspond to the $m_{\textup{CAF}}$ order parameter, so they fall to zero for certain $\alpha = \alpha_{2c}$ values. For $\lambda=1$, we recover the isotropic case studied by Oliveira \cite{oliveira}, where the critical values of $\alpha$ are $\alpha_{1c} \simeq 0.40$ and $\alpha_{2c} \simeq 0.62$. In this case $m_{\textup{CAF}}$ suffers a jump discontinuity when falling to zero. This is a signal of a first-order quantum phase transition at  $\alpha_{2c} \simeq 0.62$. We remark that for $ \alpha_{1c} < \alpha < \alpha_{2c}$ it is seen a gap where the system is in the quantum paramagnetic phase $\textup{QP}$.  The length of this gap decreases with  $\lambda$, as shown in Fig.4. Thus, for $\lambda = 0.8$,   $\alpha_{1c} \simeq 0.36$ and $\alpha_{2c} \simeq 0.52$. For $\lambda = 0.52$,   $\alpha_{1c} \simeq 0.25$ and $\alpha_{2c} \simeq 0.31$, and for $\lambda = 0.4$,   $\alpha_{1c} \simeq 0.19$ and $\alpha_{2c} \simeq 0.22$. Interestingly, for  $\lambda = 0.4$, the  $m_{\textup{CAF}}$ curve does not show a discontiuous fall, as seen for  greater values of $\lambda$, so this indicates that there is a critical value of $\lambda$ bellow which the $\alpha = \alpha_{2c}$ points are quantum critical points  of second order. 
\vskip \baselineskip 
In order to confirm the change of the behavior of the $m_{\textup{CAF}}$ curve shown in Fig.4, we plotted in Fig.5 the energy $E_{0}$ for $\lambda = 0.8$ and $\lambda = 0.4$. For $\lambda = 0.8$, we may observe that phase AF minimizes $E_{0}$ when $0 < \alpha < 0.36$, phase QP minimizes $E_{0}$ in the interval  $0.36 < \alpha < 0.52$, and phase CAF, for $\alpha > 0.52$. This is in agreement with the behavior of the curves  in Fig.4 plotted for $\lambda = 0.8$. Furthermore, it is seen a cusp at $\alpha = 0.525 \pm 0.001$, just between phases QP and CAF. This cusp,  which is a clear signal of a discontinuity of the first derivative of $E_{0}$, confirms that for this value of $\alpha$ there is a first-order quantum phase transition, corresponding to the  jump discontinuity of the  $m_{\textup{CAF}}$ curve  at $\alpha = \alpha_{2c}$. On the other hand, for $\lambda = 0.4$, the QP interval is quite reduced and no cusp is present, so  the system suffers  second-order quantum phase transitions, for $\alpha = \alpha_{1c}$ and $\alpha = \alpha_{2c}$. 
\vskip \baselineskip 
Having obtained the different values of $\alpha_{1c}$ and  $\alpha_{2c}$ for different values of $\lambda$ ensuring the minimization of $E_{0}$, we are now able to plot the frontier curves that separate the ordered phases in the $\lambda-\alpha$ plane. We show it in Fig.6, where there are  two frontiers enclosing the quantum paramagnetic phase QP. The lower frontier is of second order and divides phases AF and QP, whereas the upper one separates phases QP and CAF, in which a quantum critical point divides it into two sections. This quantum critical point is numerically located at $\lambda^{*} = 0.443(1)$ and $\alpha^{*} = 0.253(1)$, so that, for $\lambda < \lambda_{c}$, the upper frontier separating phases QP and CAF is of second order, whereas for $\lambda > \lambda^{*}$, it is of first order. The insets in Fig.4 confirm by the order paramater curves the behavior of the frontiers enclosing the QP phase. In contrast with Fig.4 of reference \cite{orlando}, the inset in Fig.6,  for $\lambda = 0.4$, exhibits not only no jump discontinuity of $m_{\textup{CAF}}$, but  the persistence of the  gap between $\alpha_{1c}$ and $\alpha_{2c}$. We verified numerically that this gap disappears only at $(\lambda,\alpha) = (0,0)$, so there is no such a first-order line dividing phases AF and CAF, such as Fig.4 of reference \cite{orlando} exhibits. 
\vskip \baselineskip 
Finally, in Fig.7 we plotted the energy $E_{0}$ as a function of $\lambda$ and $\alpha$, for two different perspectives. This figure fully describes the ground state energy, so the generated 3D surface helps us to confirm the topology of the phase diagram shown in Fig.6. For instance, we may observe a cusp partially extended along the frontier line dividing phases QP and CAF, which shows its first-order nature for $\lambda > \lambda^{*}$. 

\vskip \baselineskip 

\section{Conclusions}

We have revisited the variational study of the quantum $J_{1}-J_{1}^{'}-J_{2}$ spin-1/2 Heisenberg antiferromagnet. We obtained the phase diagram in the  $\lambda-\alpha$ plane correcting a previous result published in reference \cite{orlando}. We found numerically that there is no frontier line separating phases AF and CAF, instead, these phases are separated by two lines that meet themselves only at $(\lambda, \alpha) = (0,0)$. The lower line is of second order, whereas the upper line has two sections, the left one being of second order, and the right one of first order. These sections are  divided by a quantum critical point located at $(\lambda, \alpha) \simeq (0.443,0.253)$. We believe that these variational results may be improved by increasing the size of the plaquette, so that we may study the finite-size effect on the critical values. Nevertheless, a computational challenge must be faced, because the number of equations derived from the extremization of $\mathcal{F}$ (see Eq.(14)) will increase considerably. However, we think that efficient parallel compulational algorithms can solve this problem. 
\vskip \baselineskip 

\textbf{ACKNOWLEDGEMENT}

This work was  supported by CNPq (Brazilian Agency).

\vskip \baselineskip

\vskip \baselineskip
\vspace{3.0 cm}
\begin{figure}[htbp]
\centering
\includegraphics[height=5.0cm]{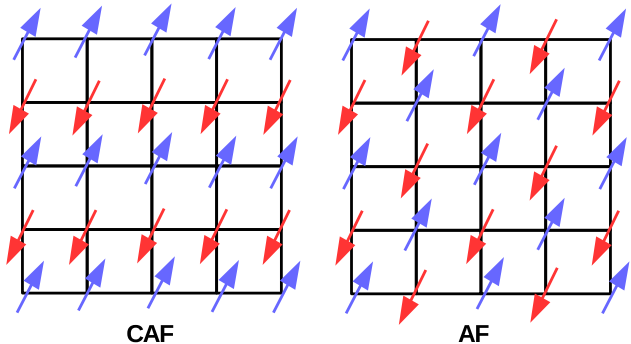}
\caption{Some spin configurations of phases CAF and AF. } 
\label{figura1}
\end{figure}

\vskip \baselineskip
\vspace{3.0 cm}
\begin{figure}[htbp]
\centering
\includegraphics[height=5.0cm]{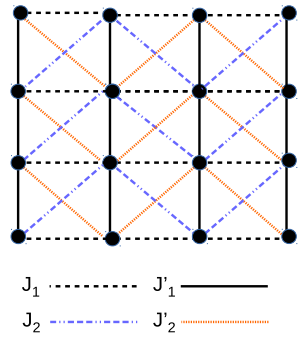}
\caption{The magnetic interactions in $\bf{[VOPO_{4}]}$ layers. See also reference \cite{tsirlin}. } 
\label{figura2}
\end{figure}

\vskip \baselineskip  

\vskip \baselineskip
\vspace{3.0 cm}
\begin{figure}[htbp]
\centering
\includegraphics[height=6.0cm]{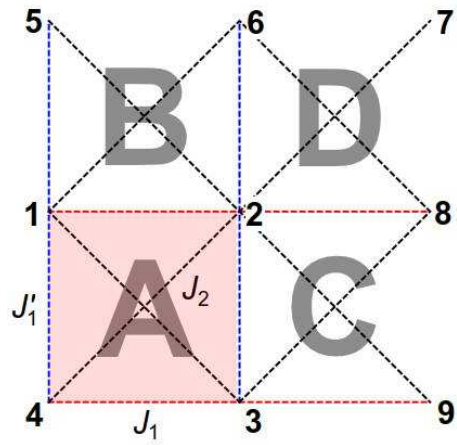}
\caption{Structure of the two-dimensional  plaquette  considered in this work. The shadowed plaquette has four spin
operators that are considered in Eq. (2). } 
\label{figura3}
\end{figure}

\vskip \baselineskip

\begin{figure}[htp]
\begin{center}
\includegraphics[height=5.0 cm]{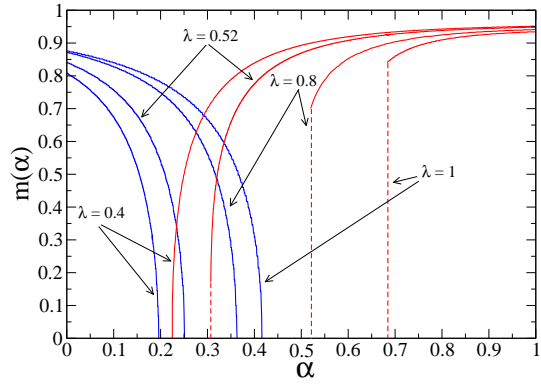}
\end{center}
\vspace{0.05cm}
\caption{\footnotesize Behavior of the ground-state order parameters $m_{\textup{AF}}$ (left curves) and $m_{\textup{CAF}}$ (right curves) of the $J_{1}-J_{1}^{'}-J_{2}$ spin-1/2 Heisenberg antiferromagnetic model as a function of the frustration parameter $\alpha$, for three values of $\lambda$. The dashed line indicates a jump discontinuity, where a first-order phase transition occurs. So, we may infere that  for $ 0.40 < \lambda < 0.52$,  there is a critical value of $\lambda$, for which  $m_{\textup{CAF}}$ changes the type of  phase transition.  }
\label{figura4} 
\end{figure}

\vskip \baselineskip
\vspace{3.0 cm}
\begin{figure}[htbp]
\centering
\includegraphics[height=5.0cm]{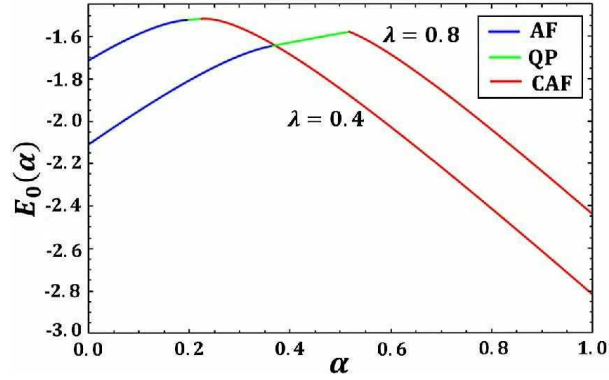}
\caption{Behavior of the ground-sate energy per plaquette (in $J_{1}$ units) of the $J_{1}-J_{1}^{'}-J_{2}$ Heisenberg model as a 
function of the frustration parameter $\alpha$, for $(a) \lambda=0.4$ and $\lambda = 0.8$. For $\lambda = 0.8$, a cusp is observed
 for $\alpha = 0.525 \pm 0.001$, at the transition between  phases QP and CAF. This is a first-order phase transition due to the discontinuous change of slope. } 
\label{figura5}
\end{figure}

\vskip \baselineskip
\begin{figure}[htp]
\begin{center}
\includegraphics[height=6.0 cm]{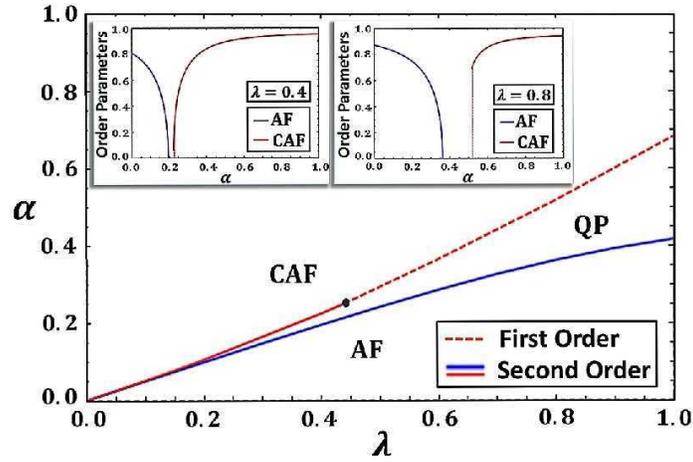}
\end{center}
\vspace{0.05cm}
\caption{\footnotesize Phase diagram of the ground state in the $\lambda$-$\alpha$ plane for the quantum
spin-1/2 $ J_{1}-J_{1}^{'}-J_{2}$ model on a square lattice, where $\alpha=J_{2}/J_{1}$ and $\lambda=J_{1}^{'}/J_{1}$ . 
The notations indicated by AF, CAF and QP corresponds the antiferromagnetic, collinear antiferromagnetic and quantum paramagnetic phases, respectively. The dashed and solid lines correspond to the first- and second-order transition lines, respectively. The upper frontier, which limits the CAF phase, consists of a second-order and a first-order section divided by a quantum critical point represented by the black point. This can  be shown by the  inner figures, in which  the CAF order parameter   (the curve on the right in both insets) suffers a jump discontinuity when falling to zero, for $\lambda = 0.8$, whereas, for $\lambda = 0.4$, it falls continuously to zero (see also figure 5).  }
\label{figura6} 
\end{figure}

\vskip \baselineskip
\vspace{3.0 cm}
\begin{figure}[htbp]
\centering
\includegraphics[height=4.5cm]{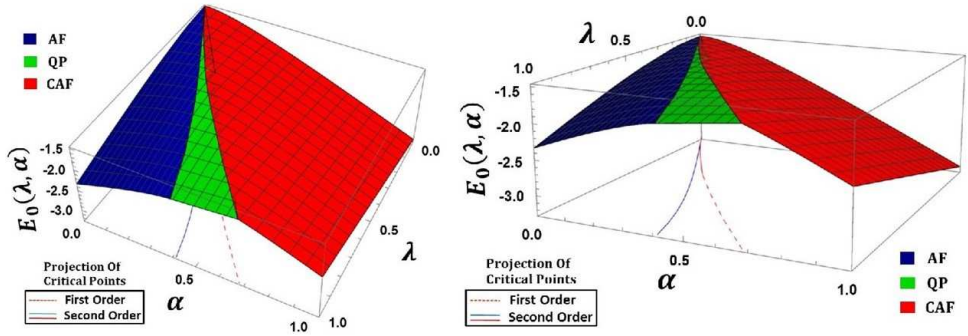}
\caption{The Energy surface $E_{0}$  saw by two different perspectives that complements the information of Fig.5. Here we have a complete view on  which phase minimizes $E_{0}$, according to the region of the $\lambda - \alpha$ plane. The figure on the right may betterly exhibit the discontinuity of the first derivative of $E_{0}$ along some section of the frontier dividing the QP and CAF phases. It signals its partial first-order nature, as observed after the black critical point shown in Fig.4. } 
\label{figura7}
\end{figure}

\end{document}